\documentclass[aps,prb,twocolumn]{revtex4}
\usepackage{graphicx}

\begin{document}

\title{Specific Heat Study of the Magnetic Superconductor HoNi$_{2}$B$_{2}$C}
\author{Tuson Park}
\altaffiliation {Present address: Los Alamos National Laboratory, Los Alamos, NM 87545, USA} \email{tuson@lanl.gov}
\author{M. B. Salamon}
\affiliation{Department of Physics and Material Research Laboratory, University of\\
Illinois at Urbana-Champaign, IL 61801, USA}
\author{Eun Mi Choi, Heon Jung Kim, and Sung-Ik Lee}
\affiliation{National Creative Research Initiative Center for Superconductivity and
Department of Physics, Pohang University of Science and Technology, Pohang\\
790-784, Republic of Korea}
\date{\today}

\begin{abstract}
The complex magnetic transitions and superconductivity of HoNi$_{2}$B$_{2}$C
were studied via the dependence of the heat capacity on temperature and
in-plane field angle. We provide an extended, comprehensive magnetic phase
diagram for $B\parallel \lbrack 100]$ and $B\parallel \lbrack 110]$ based on
the thermodynamic measurements. Three magnetic transitions and the
superconducting transition were clearly observed. The 5.2~K transition ($%
T_{N}$) shows a hysteresis with temperature, indicating the first order
nature of the transition at $B=0$~T. The 6~K transition ($T_{M}$), namely
the onset of the long-range ordering, displays a dramatic in-plane
anisotropy: $T_{M}$ increases with increasing magnetic field for $B\parallel
\lbrack 100]$ while it decreases with increasing field for $B\parallel
\lbrack 110]$. The anomalous anisotropy in $T_{M}$ indicates that the
transition is related to the a-axis spiral structure. The 5.5~K transition ($%
T^{\ast }$) shows similar behavior to the 5.2~K transition, i.e., a small
in-plane anisotropy and scaling with Ising model. This last transition is
ascribed to the change from $a^{\ast }$ dominant phase to $c^{\ast }$
dominant phase.
\end{abstract}

\maketitle

\section{Introduction}

An understanding of the interplay between magnetism and superconductivity
has been an area of intensive research because of their seemingly
antagonistic tendencies.\cite{fischer90,bulaevskii85} The limited
availability of suitable examples and the low magnetic transition
temperatures of those that exist have made the studies very difficult
experimentally. The recently found RNi$_{2}$B$_{2}$C family where R is
rare-earth element offered a new venue because the magnetic transitions ($%
T_{M}$) occur in an easily accessible temperature range with a variation of $%
T_{M}/T_{C}$ ranging from 1.75 for Dy to 0.14 for Tm. Further, high-quality
single crystals have become available. Among the magnetic members, HoNi$_{2}$%
B$_{2}$C is particularly interesting because its complex magnetic phases are
observed to coexist with superconductivity. Neutron scattering revealed
three types of magnetic order.\cite{goldman94,grigereit94,kreyssig97} A
commensurate antiferromagnetic structure is formed with $q=c^{\ast }$, i.e.
(0~0~1) below 6~K in which the spins order ferromagnetically within the $a-b$
plane and antiferromagnetically along the $c-$axis. Two incommensurate
structures coexist over a finite temperature range, i.e., 5~K $\leq T\leq 6$%
~K: the $c^{\ast }$ structure with $q=(0~0~0.915)$ where the
ferromagnetically aligned spins are rotated by $165^{\circ }$ and the $%
a^{\ast }$ structure with $q=(0.585~0~0)$ for which the detailed structure
is still unknown.

HoNi$_{2}$B$_{2}$C exhibits a near-reentrant superconductivity, i.e.,
reentrant resistive behavior in a small magnetic field, due to competition
between superconductivity and exchange-coupled antiferromagnetic order.\cite%
{eisaki94} In a theoretical analysis of the interplay, the onset of the $c-$%
axis incommensurate state has been shown to suppress superconductivity,
leading to the near-reentrant behavior.\cite{amici98,amici00} Kreyssig et
al., however, found that in Y-doped quasi-quartenary compound, Ho$_{1-x}$Y$%
_{x}$Ni$_{2}$B$_{2}$C, only the incommensurate $a-$axis feature remains in
the same temperature range as the near-reentrant behavior while the $c-$axis
spiral structure exists over a much wider temperature range, indicating that
it is the $a^{\ast }$-structure connected with Fermi surface nesting that
enhances the pair-breaking effect.\cite{kreyssig97} For a further
understanding of the interplay, it is necessary to study the nature of the
magnetic transitions and to establish the $B-T$ phase diagram of HoNi$_{2}$B$%
_{2}$C.

The magnetic phase diagram of HoNi$_{2}$B$_{2}$C has been studied
extensively in the context of the interplay between magnetism and
superconductivity, mostly below $B=1$~T.\cite{dumar96} Zero-field specific
heat\cite{canfield94} and other surface sensitive measurements\cite%
{rathnayaka96,elhagary98} indicate that there are three magnetic transitions
($T_{N}$, $T^{\ast }$, $T_{M}$), while other specific heat data\cite%
{lin95,choi01} show only two transitions ($T_{N}$, $T_{M}$), where $%
T_{N}=5.2 $~K is the Neel temperature, $T_{M}=6$~K is the onset of a
long-range magnetic ordering, and $T^{\ast }=5.5$~K is ascribed to some
change of the oscillatory magnetic state. It has been suggested that the low
temperature physical properties of polycrystalline HoNi$_{2}$B$_{2}$C depend
on thermal treatment as well as chemical composition while the properties of
single crystals are relatively less affected.\cite{dertinger01,wagner99}
Even though majority of measurements indicate the presence of $T^{\ast }$ at
5.5~K, thermodynamic measurements on single crystals that show the magnetic
transition to be a truly bulk property are rare.

In this paper, we report specific heat measurements of single crystal HoNi$%
_{2}$B$_{2}$C as a function of temperature and magnetic field to provide an
extended (up to 6~T), comprehensive magnetic phase diagram for $B\parallel
\lbrack 100]$ and $B\parallel \lbrack 110]$. Three magnetic transitions are
clearly resolved, confirming that $T^{\ast }$ is a bulk magnetic transition.
The 5.2~K transition ($T_{N}$) shows hysteresis with temperature, indicating
a first order antiferromagnetic transition. The 6~K transition ($T_{M}$),
namely the onset of long-range order, displays a dramatic in-plane
anisotropy: $T_{M}$ increases with increasing magnetic field for $B\parallel
\lbrack 100]$ while it decreases with increasing field for $B\parallel
\lbrack 110]$. The anomalous anisotropy in $T_{M}$ indicates that the
transition is related to the $a$-axis spiral structure. The 5.5~K transition
($T^{\ast }$) shows a small in-plane anisotropy and was ascribed to the
transition from $a^{\ast }$-dominant phase to $c^{\ast }$ dominant phase.
The specific heat jump related to the superconducting transition was
observed at 8~K.

\section{Superconductivity}

We have used two single crystals of HoNi$_{2}$B$_{2}$C, labeled sample A and
sample B for the heat capacity measurements. Both samples were from a same
batch. A high temperature flux method was used to grow the sample using Ni$%
_{2}$B as a solvent. The details are described elsewhere.\cite{cho95}

\begin{figure}[tbp]
\includegraphics[width=7cm,clip]{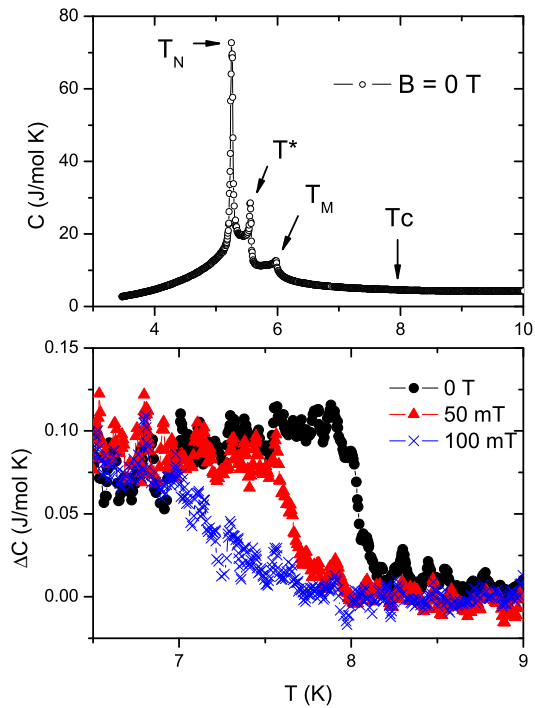}\centering
\caption{Top panel: the specific heat of sample A in zero field. Bottom
panel: the specific heat discontinuity, $\Delta C(B,T)=C(B,T)-C(0.2T,T)$, in
the vicinity of $T_{c}$ for $B=0,50,100$~mT}
\end{figure}
The heat capacity measured by ac calorimetry \cite{tuson02} was converted to
absolute values using literature data.\cite{lin95} The upper panel of Fig.~1
shows the specific heat of HoNi$_{2}$B$_{2}$C vs. temperature at zero field.
The three sharp peaks mark the magnetic transitions noted above. The first
transition at 6~K is marked as $T_{M}$, the second at 5.5~K, as $T^{\ast }$,
and the third at, 5.2~K as $T_{N}$. There is also a specific heat
discontinuity around 8~K related to superconducting transition but it is too
small to be seen on the same scale.

The bottom panel shows the specific heat discontinuity related to the
superconducting transition in the vicinity of $T_{c}$. The circles describe
the specific heat difference, $\Delta C=C-C_{n}$, for $B=0$~T, the triangles
for $B=50$~mT, and the crosses for $B=100$~mT, where $C_{n}$ is the specific
heat in normal state. Data taken at 0.2~T were used as the normal state
background between 7~K and 9~K because the magnetic contribution is
unaffected at low fields and the superconductivity is suppressed in that
temperature range. In zero field, the transition from normal to
superconducting state occurs at 8.04~K with a narrow transition width ($%
\Delta T/T_{c}\leq 0.04$). The transition temperature was defined as the
midpoint of the transition, which is essentially equal to that found from
entropy-conserving method. As the magnetic field increases, the $T_{c}$
decreases at the rate -8.3 K/T and the transition width becomes broadened.

The specific heat jump in zero field is about 110 mJ/mol$\cdot $K. If we use
the BCS relation $\Delta C/\gamma T_{c}=1.43$, we obtain $\gamma =9.6$~mJ/mol%
$\cdot $ K$^{2}$, small compared to that of non-magnetic counterpart, i.e.
18~mJ/mol$\cdot $K$^{2}$ for Lu(Y)Ni$_{2}$B$_{2}$C, possibly indicating a
lower density of states, $N(E_{F})$, in HoNi$_{2}$B$_{2}$C.\cite{carter95} A
spectroscopic study, however, found that the density of states hardly
changes within the borocarbide series (RNi$_{2}$B$_{2}$C).\cite{pellegrin95}
Recently, El-Hagary et al. found a common correlation between the specific
heat jump $\Delta C$ and the transition temperature $T_{c}$ among magnetic
borocarbide superconductors, i.e., $\Delta C\propto T_{c}^{2}$,\cite{elhagary00} which indicates the importance of the magnetic pair breaking.
According to Abrikosov-Gor'kov theory (AG), the exchange interaction between
electrons and magnetic impurity atoms leads to nonconservation of the
electron spin, affecting the formation of Cooper pairs. Assuming $\gamma
\sim 18$~mJ/mol$\cdot $K$^{2}$ (same as that of Y or Lu based borocarbide)%
\cite{michor95}, the ratio $\Delta C/\gamma T_{c}$ becomes 0.76. If we
assume a 50~\% suppression of $T_{c}$ from the nonmagnetic value of 16 K,
the corresponding AG prediction is $\Delta C/\gamma T_{c}\sim 1$.

It can be speculated that the origin of the small specific heat discontinuity
results from anisotropic superconductivity or a multi-band superconductivity
as is the case for the two-gap superconductor MgB$_{2}$.\cite%
{haas01,mishonov02,bouquet02,tuson02,yang01} Indeed, there is compelling evidence that the non-magnetic members of the borocarbides, Lu(Y)Ni$_{2}$B$_{2}$C,
are highly anisotropic or probably nodal superconductors where there exist
gap zeros on the Fermi surface.\cite{nohara97,boaknin01,tuson03} The
positive curvature in the upper critical field $H_{c2}$ of Lu(Y)Ni$_{2}$B$%
_{2}$C close to $T_{c}$ and the temperature dependence of $H_{c2}$ were
successfully explained by using an effective two-band model.\cite{shulga98}
However, the gap anisotropy or multi-band feature is probably irrelevant to
the anomalous value of $\Delta C/\gamma T_{c}$ in HoNi$_{2}$B$_{2}$C because
the reported thermodynamic ratio ($=2.3$) of LuNi$_{2}$B$_{2}$C\cite%
{nohara97} is much larger than the weak-coupling BCS value (=1.43) as well
as that of HoNi$_{2}$B$_{2}$C~($=0.76$).

\section{Magnetic phase transitions}

\subsection{Magnetic transition at 5.2~K ($T_{N}$)}

\begin{figure}[tbp]
\includegraphics[width=7cm,clip]{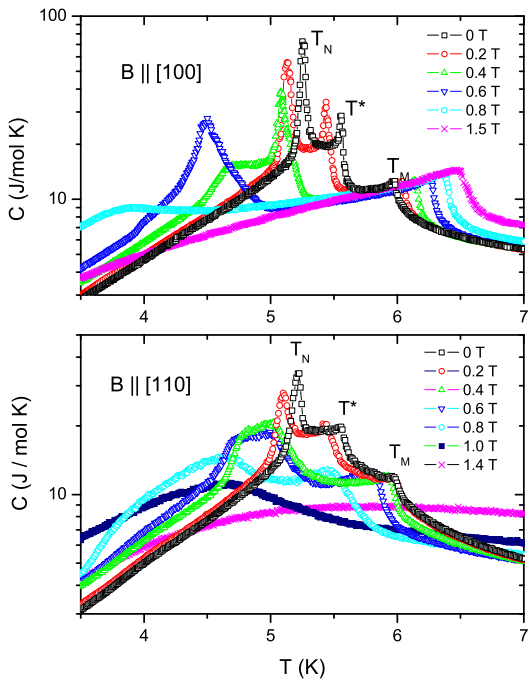}\centering
\caption{The top panel describes the specific heat of sample A as a function
of temperature in several constant fields along [100] and the bottom panel
represents the specific heat of sample B in fields along [110]. Both plots
are on a semi-log scale}
\end{figure}
The top panel of Fig.~2 shows a semi-log plot of the specific heat of HoNi$%
_{2}$B$_{2}$C (sample A) as a function of temperature at several magnetic
fields $B\parallel \lbrack 100]$. The bottom panel of Fig.~2 describes the
specific heat of sample B for $B\parallel \lbrack 110]$. The $c$-axis
commensurate antiferromagnetic (AF) transition, labeled $T_{N}$, is lowered
with increasing field for both field directions. When field is higher than
0.4~T, the AF peak becomes severely broadened, making the data difficult to
interpret. The top panel of Fig.~3 shows the $B-T$ phase diagram for this
transition to a N\'{e}el state. The critical temperatures $T_{N}$ from $C$
vs. $T$ in constant fields (\textbf{squares}) and the critical fields $B_{N}$
from the isothermal $C$ vs. $B$ at constant temperatures (\textbf{circles};
see Figs.~4 and 5) were plotted for both field directions. There is
negligible in-plane $B_{N}$ anisotropy between [100] and [110] directions.
The dashed line is the least-squares fit of $B_{N}=A(1-T/T_{N})^{0.5}$ in
the vicinity of $T_{N}$, showing Ising-like behavior. \cite{fisher60,keen66}
Experimental data were explained well with $A=1.41\pm 0.01$~T. 
\begin{figure}[tbp]
\includegraphics[width=7cm,clip]{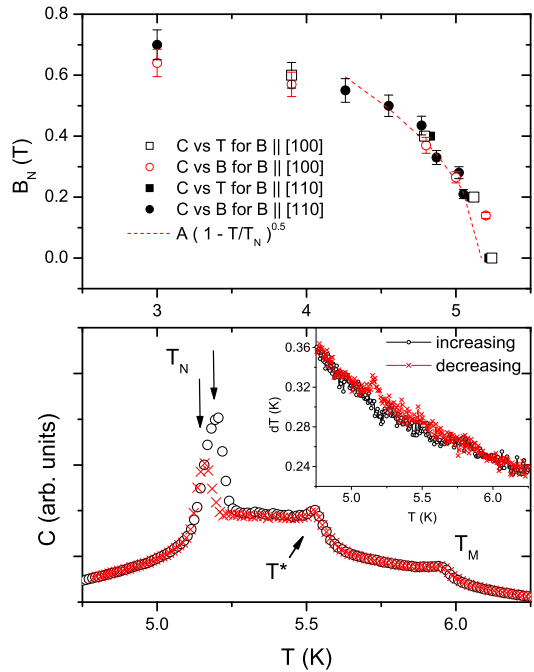}\centering
\caption{The top panel describes $B_{N}-T$ phase diagram for $B\parallel
\lbrack 100]$ (\textbf{open symbols}) and $B\parallel \lbrack 110]$ (\textbf{%
solid symbols}). The circles represent the data from isothermal $C$ vs. $B$
and the squares, from $C$ vs $T$ in several fields. The dashed line
describes the least square fit of $A(1-T/T^{\ast })^{0.5}$ with $A=1.41\pm
0.01$~T. The bottom panel shows the specific heat of sample B as a function
of temperature in zero field. The circles represent the data with increasing
temperature and the crosses, with decreasing temperature. Inset: the
temperature offset of the sample as a function of temperature.}
\end{figure}

The bottom panel of Fig.~3 shows the hysteresis of the specific heat of
sample B as a function of temperature. The circles describe the data with
increasing temperature while the crosses, with decreasing temperature. There
is no hysteresis in either the 6~K transition $T_{M}$ nor the 5.5~K
transition $T^{\ast }$. In the AF transition, however, there is a clear
hysteresis where the AF peak was moved from 5.21~K with increasing
temperature to 5.16~K with decreasing temperature. The inset of the bottom
panel shows the $dc$ temperature offset of the sample as a function of
temperature. Depending on the direction of the temperature sweep, there
occurs a sudden jump (or drop) in the dc offset at $T_{N}$, a consequence of
the latent heat associated with the transition. The rectangle-shaped
hysteresis adds additional strong evidence that the $T_{N}$ transition is
first order. This finding is consistent with the observation of
magnetoelastic effects in HoNi$_{2}$B$_{2}$C, where the length of the unit
cell in [110] direction is shortened by about 0.19~\% compared to its length
in $[\bar{1}10]$ direction below $T_{N}$.\cite{kreyssig99} Despite the
square-root behavior evident in Fig. 3,\ the first-order nature of the
magnetic AF transition disagrees with the Ising model.\cite{fisher60} We
note that the peak intensity is smaller for the specific heat with
decreasing temperature. That could be an artifact from ac calorimetry
because the temperature response of the sample is not an ideal triangular
shape, rather a distorted one when it goes through a phase transition with
latent heat.\cite{garnier72} 
\begin{figure}[tbp]
\includegraphics[width=7cm,clip]{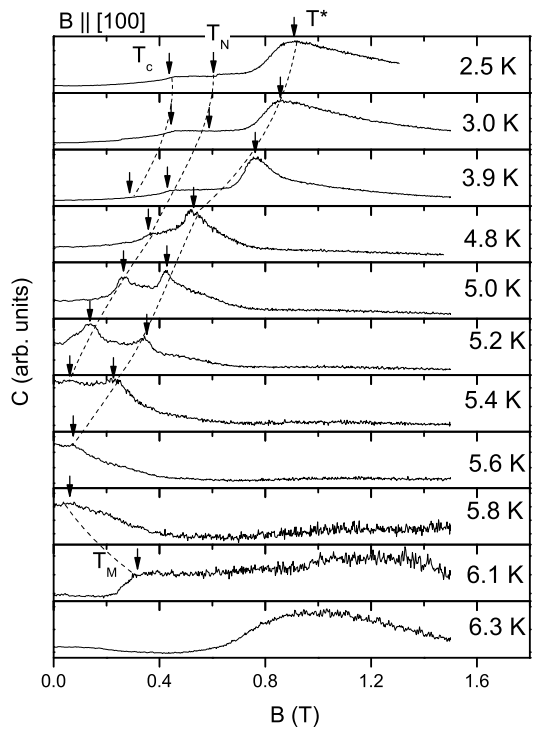}\centering
\caption{Isothermal specific heat of sample B as a function of magnetic
field at several temperatures for $B\parallel \lbrack 100]$. Prominent peaks
and shoulders were marked by arrows. The peaks and shoulders are more
evident on an expanded scale. The transition from superconducting state to
normal state is marked as $T_{c}$. The dotted lines are guides to the eye.}
\end{figure}

\begin{figure}[tbp]
\includegraphics[width=7cm,clip]{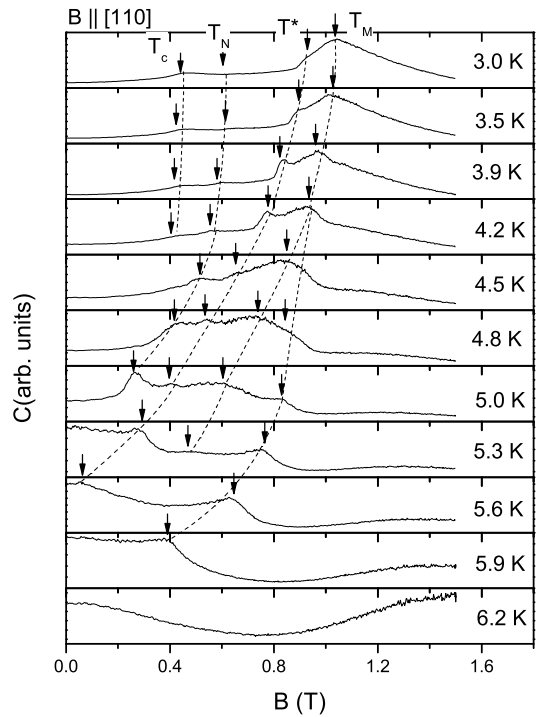}\centering
\caption{Isothermal specific heat of sample B as a function of magnetic
field at several temperatures for $B\parallel \lbrack 110]$. Prominent peaks
and shoulders were marked by arrows. The peaks (or shoulders)at each
temperature were determined on an expanded scale. The transition from
superconducting state to normal state is marked as $T_{c}$. The dotted lines
are guides to the eye.}
\end{figure}

\subsection{The 5.5~K magnetic transition ($T^{*}$)}

The 5.5~K magnetic transition ($T^{\ast }$) is as sharp as the AF
transition, for sample A at least, indicating it could also be a first order
phase transition (see Fig.~2). The magnetic field dependence of the specific
heat is also similar to the $T_{N}$ counterpart, i.e. the critical
temperature decreases with increasing magnetic field. Initially, the peak
intensity at $T^{\ast }$ becomes stronger with increasing field while that
of the AF transition decreases monotonically, transferring some of its
entropy to the $T^{\ast }$ transition. Fig.~6 shows the $B^{\ast }-T$ phase
diagram both for $B\parallel \lbrack 100]$ (\textbf{open}) and $B\parallel
\lbrack 110]$ (\textbf{solid}), where $B^{\ast }$ is the critical field
corresponding to the $T^{\ast }$ transition. The in-plane $B^{\ast }$
anisotropy between [100] and [110] directions is small, similar to the $B_{N}
$ transition. The $B^{\ast }$ temperature dependence near $T^{\ast }$ was
explored in terms of the antiferromagnetic theory by Fisher,\cite{fisher60}
as was done in the $T_{N}$ transition. The suppression of the critical
temperature was explained reasonably well by $B^{\ast }=A(1-T/T^{\ast
})^{0.5}$ with $A=1.39\pm 0.01$~T (\textbf{dashed line}), a value similar to
that of the $T_{N}$ analysis.
\begin{figure}[tbp]
\includegraphics[width=7cm,clip]{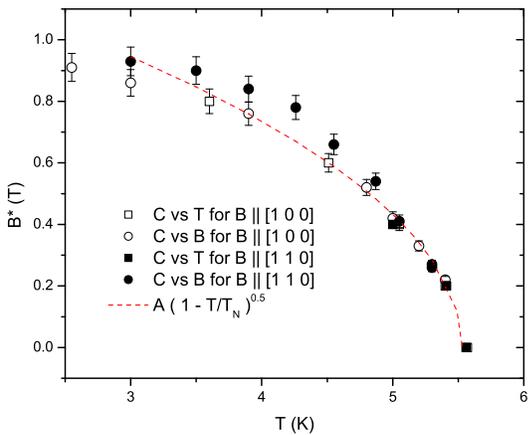}\centering
\caption{$B^{\ast }-T$ phase diagram for both field directions: open symbols
describe the data for $B\parallel \lbrack 100]$ and solid symbols, for $%
B\parallel \lbrack 100]$. The squares represent the data obtained from $C$
vs. $T$ while the circles, from isothermal $C$ vs. $B$ data. The dashed line
describes the least square fit of $A(1-T/T^{\ast })^{0.5}$ with $A=1.39\pm
0.01$~T.}
\end{figure}

What could be responsible for the 5.5~K transition? Neutron scattering
showed that all three magnetic structures, i.e. $c$-axis commensurate AF
magnetic structure, $c$-axis AF helical structure, and $a$-axis
incommensurate structure, coexist between 5~K and 6~K.\cite{lynn97,goldman94}
In the previous section, we argued that the 5.2~K transition is due to the
magnetic transition to the $c$-axis commensurate AF structure ($T_{N}$) and
is a first order phase transition. All the similarities between the 5.5~K
transition and the 5.2~K transition point toward associating $T^{\ast }$
with the $c$-axis incommensurate AF structure. Then, the next question is
\textquotedblleft Is it a first order phase transition?\textquotedblright\
The specific heat at zero field did not show any noticeable hysteresis with
temperature at $T^{\ast }$ (see Fig.~3). In the dc temperature offset (see
the inset of Fig.~3), however, there may be a small hysteresis at $T^{\ast }$%
, but the feature is well within the scattered data. Coexistence of the
phases also suggests that it is 1st order. At this point, it is not clear if
the 5.5~K magnetic transition is a first order phase transition or not.

\subsection{The 6~K magnetic transition ($T_{M}$)}

The 6~K magnetic transition ($T_{M}$) is known to be the onset of a
long-range magnetic order. The precise nature of the $T_{M}$ transition is
not well characterized, although recent work \cite{detlefs00} argues that it
represents the onset of the $a^{\ast }$ modulated phase. In this section, we
present the specific heat as a function of temperature both for $B\parallel
\lbrack 100]$ and $B\parallel \lbrack 110]$ directions and confirm that the
6~K transition is due to the onset of the $a$-axis incommensurate magnetic
structure.

Fig.~2 shows a dramatic difference in the critical temperature $T_{M}$ with
magnetic field directions. For $B\parallel \lbrack 110]$, the critical
temperature scarcely changes with increasing magnetic field below 0.5~T in
agreement with Detlefs et al. \cite{detlefs00} and Du Mar, et al.\cite{dumar96}
In higher fields, it decreases rapidly as is expected for an
antiferromagnetic transition and the transition shape becomes broadened,
making it hard to interpret. For $B\parallel \lbrack 100]$, the critical
temperature increases with increasing magnetic field\cite{dumar96} while
keeping its transition shape. The anomalous in-plane anisotropy of $T_{M}$
strongly suggests that the origin of this transition is very different from
the other two magnetic transitions ($T_{N}$, $T^{\ast }$) and is, therefore,
due to the onset of the $a-$axis incommensurate magnetic structure. Detlefs
et al.\cite{detlefs00} recently hypothesized that the $a^{\ast }$ magnetic
structure is related to the Fermi surface (FS) nesting along [100]
direction. Since a distortion along [110] direction is more disruptive to
the nesting feature than a distortion along [100] direction, it results in
the absence of the $a^{\ast }$ phase for HoNi$_{2}$B$_{2}$C and DyNi$_{2}$B$%
_{2}$C at low temperature and low magnetic field where local magnetic
moments are aligned along [110] directions. If the $a^{\ast }$ phase is the
ground state for 5.5~K $\leq T\leq $6.0~K in HoNi$_{2}$B$_{2}$C, a magnetic
field applied along [110] will disrupt the magnetic phase while the field
along [100] will not. The anomalous magnetic field dependence of specific
heat is consistent with this scenario. The enhancement of the $a^{\ast }$
phase for $B\parallel \lbrack 100]$, however, is beyond this explanation.

\begin{figure}[tbp]
\includegraphics[width=7cm,clip]{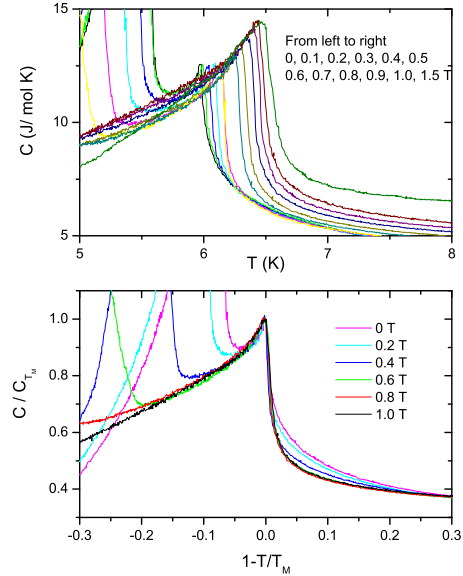}\centering
\caption{The top panel blows up the specific heat of sample A vs.
temperature near $T_{M}$ in several magnetic fields when $B\parallel \lbrack
100]$: from left to right, 0.1, 0.2, 0.3, 0.4, 0.5, 0.6, 0.7, 0.8, 0.9, 1.0,
1.5~T. The bottom panel shows a scaling behavior of $T^{\ast }$ transition
where x-axis is the reduced temperature $1-T/T_{M}$ and y-axis is the
normalized specific heat $C/C_{T_{M}}$ where $C_{T_{M}}$ is the specific
heat at $T_{M}$. For clarity, selective data were shown.}
\end{figure}
The top panel of Fig.~7 shows the specific heat data around $T_{M}$ in
several constant magnetic fields from 0~T to 1.5~T for $B\parallel \lbrack
100]$. For clarity, selective data are shown in the bottom panel, where $x$%
-axis is the reduced temperature $(1-T/T_{M})$ and $y$-axis is the specific
heat divided by the specific heat at $T_{M}$, i.e. $C/C_{T_{M}}$. All of
them collapse onto each other, showing a scaling behavior. The specific heat
near a critical point diverges logarithmically in the two-dimensional Ising
model while it diverges more strongly, as a power law in the
three-dimensional Ising model.\cite{kadanoff67} Fig.~8 describes the
zero-field magnetic specific heat as a function of the reduced temperature $%
\tau (=1-T/T_{M})$ on a semi-log scale. The magnetic specific heat was
obtained by subtracting the lattice and electronic contributions: $%
C_{M}=C-\gamma T-\beta T^{3}$ where we used $\gamma $ and $\beta $ values of
TmNi$_{2}$B$_{2}$C.\cite{movshovich94} The higher temperature side of $T_{M}$
was analyzed in terms of 2D and 3D Ising model. The dashed line is the best
fit of 3D Ising model with a functional form of $A~|\tau |^{-0.1}+B$. The
solid line is from 2D Ising model of $-C~log|\tau |+D$. Both 2D and 3D Ising
models can explain the data over two decades of temperature range, i.e. $%
3\times 10^{-3}<\tau <3\times 10^{-1}$. When temperature is close enough to $%
T_{M}(=5.98)$~K, however, the specific heat data can be explained better
with the logarithmic function than the power-law dependence. The logarithmic
singularity may be interpreted as a manifestation of two-dimensional spin
structure in HoNi$_{2}$B$_{2}$C where the Ho$^{3+}$ local moment is confined
to the Ho-C basal plane for temperatures below 100~K.\cite{lynn97} We note
that the value of the coefficient of the logarithmic term, $C=3$~J/mol$\cdot 
$~K, is of the order of magnitude found in the exact theory of 2D Ising
antiferromagnets.\cite{onsager44} The lower temperature side of $T_{M}$ is
severely contaminated by the adjacent magnetic transition $T^{\ast }$,
rendering analysis difficult.
\begin{figure}[tbp]
\includegraphics[width=7cm,clip]{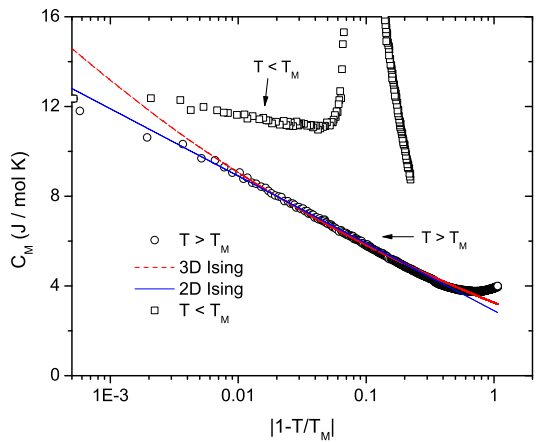}\centering
\caption{Magnetic specific heat of sample A as a function of the reduced
temperature $\protect\tau (=1-T/T_{M})$ in zero field on a semi-log scale.
The dashed line is the least square fit of 3D Ising model, i.e. $A~|\protect%
\tau |^{-0.1}+B$ with $A=9.92$ J/mol~K and $B=-6.68$ J/mol K. The solid line
represents the least square fit of 2D Ising model, i.e. $-C~log|\protect\tau %
|+D$ with $C=3.0$ J/mol~K and $D=2.9$ J/mol K.}
\end{figure}

\subsection{Magnetic phase diagram}

It has long been known that there is an extreme magnetic anisotropy
associated with the crystalline electric field (CEF) splitting of the Hund's
rule ground state for the magnetic members of the borocarbide family RNi$%
_{2} $B$_{2}$C (R=Er, Tb, Ho, Dy).\cite{cho95,cho96} The R$^{3+}$ local
moment is confined to the R-C basal plane for temperatures below roughly
100~K, i.e. temperatures well above the magnetic ordering temperatures. In
addition, a strong in-plane anisotropy has been observed, leading to the
local moments essentially being confined to either the [100] (R~=~Er, Tb) or
[110] (R~=~Ho, Dy).\cite{canfield97} Therefore, it became necessary to study
physical properties as a function of magnetic field direction.

Fig.~9 summarizes the magnetic phase diagram of HoNi$_{2}$B$_{2}$C both for $%
B\parallel \lbrack 110]$ (\textbf{upper panel}) and for $B\parallel \lbrack
100]$ (\textbf{lower panel}) directions based on specific heat as a function
of temperature and magnetic field. The 5.2~K transition, $T_{N}$, is due to
the commensurate $c-$axis AF structure and found to be a first order phase
transition. The critical field $B_{N}$ (\textbf{triangles}) shows little
in-plane anisotropy. The 5.5~K transition, $T^{\ast }$, is due to the $c$%
-axis oscillatory magnetic structure. The in-plane anisotropy of $B^{\ast }$
(\textbf{circles}) is small and the nature of the transition is still
controversial. The 6~K transition, $T_{M}$, is ascribed to the onset of the $%
a$-axis oscillatory magnetic structure. The in-plane anisotropy of $B_{M}$ (%
\textbf{squares}) is very surprising: the critical temperature decreases
with increasing magnetic field for $B\parallel \lbrack 110]$ while it
increases, for $B\parallel \lbrack 100]$. The critical temperature $T_{M}$
initially increases with magnetic field at a rate of 0.48~K/T, then it
starts to bend around at about 1.5~T and decreases (see the inset of the
bottom panel of Fig.~9). The decrease of the critical temperature is
expected for antiferromagnet.\cite{fisher60} The increase, however, is
difficult to understand. 
\begin{figure}[tbp]
\includegraphics[width=7cm,clip]{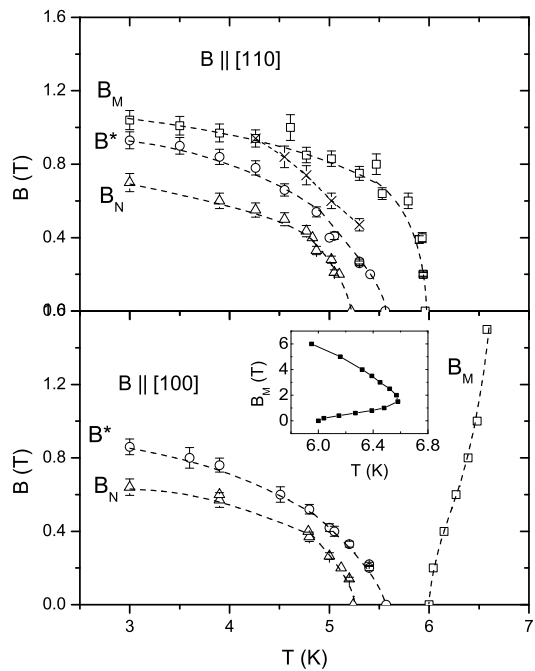}\centering
\caption{$B-T$ magnetic phase diagram both for $B\parallel \lbrack 110]$ (%
\textbf{upper panel}) and for $B\parallel \lbrack 100]$ (\textbf{lower panel 
}) directions. The triangles describe the $c-$axis commensurate AF transition,
the circles the $c-$axis incommensurate AF transition, and the squares the 
$a-$axis incommensurate transition. The crosses in the upper panel represents an additional phase line for $B \parallel \lbrack 110]$, most readily seen as a shoulder in the 4.8~K and 5.0~K in Fig.~5. Dashed lines are guides to the eye. The
insert in the bottom panel is the full $B_{M}-T$ phase diagram up to 6 T for 
$B\parallel \lbrack 100]$.}
\end{figure}

The magnetic phase diagram built from specific heat data is consistent with
the viewpoint of three distinct magnetic transitions between $5K\leq T\leq 6K
$ at zero field.\cite{dumar96,rathnayaka96} At low fields, the critical temperature $T_{M}$ scarcely changes with magnetic field for $B\parallel [110]$ in agreement with Du Mar et al.\cite{dumar96} while the increase of $T_{M}$ for $B\parallel [100]$ is more evident, probably due to high-quality sigle crystal used in this study. In the high-field regime, our data show a dramatic difference in $T_{M}$ between the two field directions, i.e., $B\parallel \lbrack 100]$ and $B\parallel \lbrack 110]$ as is evident in Fig.~9 while previous studies drew a conclusion of similar behavior between the two directions. For $B\parallel [110]$, Ref~9 shows only two transitions below 5~K, i.e., $T_{N}$ and $T^{*}$. In contrast, our data clearly show three transitions at the same temperature range. A close examination reveals that the $T^{*}$ marked transition below 5~K in Ref~9 corresponds to $T_{M}$ in our phase diagram, indicating that the real $T^{*}$ transition is missing in that phase diagram. 

There has been a speculation of a fourth magnetic phase based on resistance measurements where there occurs a slope change at 3.8~K for $B\parallel \lbrack 100]$.\cite{rathnayaka96} In our bulk measurement, the additional feature was not observed, suggesting that the 3.8~K feature is an extrinsic property
sensitive to the surface state. Instead, we found an additional phase line
for $B\parallel \lbrack 110]$, most readily seen as a shoulder in the 4.8~K
and 5.0~K data in Fig.~5. It separates the $B_{M}$ and the $B^{\ast }$ lines
and is merged to $B_{M}$ at 4~K (\textbf{crosses} in the upper panel of Fig.~9). The origin of this feature has yet to be elucidated.

\section{Magnetic field-angle heat capacity}

In non-magnetic borocarbides, the low temperature heat capacity directly
measures the electronic density of states.\cite{tuson03} In magnetic
systems, however, the magnetic specific heat dominates. For example, the
electronic and lattice components of the specific heat account for less than
5\% of the total below 8~K in HoNi$_{2}$B$_{2}$C. Consequently, in HoNi$_{2}$%
B$_{2}$C, the field-angle dependent heat capacity used to investigate the
superconducting gap nature in the non-magnetic systems, here mainly explores
the magnetic structure. In this section, we present the in-plane field-angle
heat capacity of the Ho-based borocarbide and discuss the data based on the
phase diagram built in the previous sections.

Fig.~10 shows the low-field heat capacity vs field angle measured with
respect to the $a-$axis of HoNi$_{2}$B$_{2}$C at several different
temperatures. The periodicity of the peaks is 90 degrees at all
temperatures, indicating that the peak in the specific heat is simply due to
the in-plane anisotropy between [100] and [110] directions. Canfield et al.
observed magnetization modulation as a function of magnetic field angle at
2~K.\cite{canfield97} The oscillation feature was then interpreted in terms
of metamagnetic states, i.e., $(\uparrow \downarrow )$ for $B\leq B_{c1}$, $%
(\uparrow \uparrow \downarrow )$ for $B_{c1}\leq B\leq B_{c2}$, $(\uparrow
\uparrow \rightarrow )$ for $B_{c2}\leq B\leq B_{c3}$ and $(\uparrow
\uparrow \uparrow )$ for $B\geq B_{c3}$. Here the arrow $\uparrow $ is a
moment along the [110] axis, $\downarrow $ is a moment along the $[\bar{1}\bar{1}0]
$, and $\rightarrow $ is a moment along the $[1\bar{1}0]$ axis. In our
magnetic phase diagram (Fig. 9), $B_{c1}$ corresponds to $B_{N}$, $B_{c2}$
to $B^{\ast }$, and $B_{c3}$ to $B_{M}$ respectively.
\begin{figure}[tbp]
\includegraphics[width=7cm,clip]{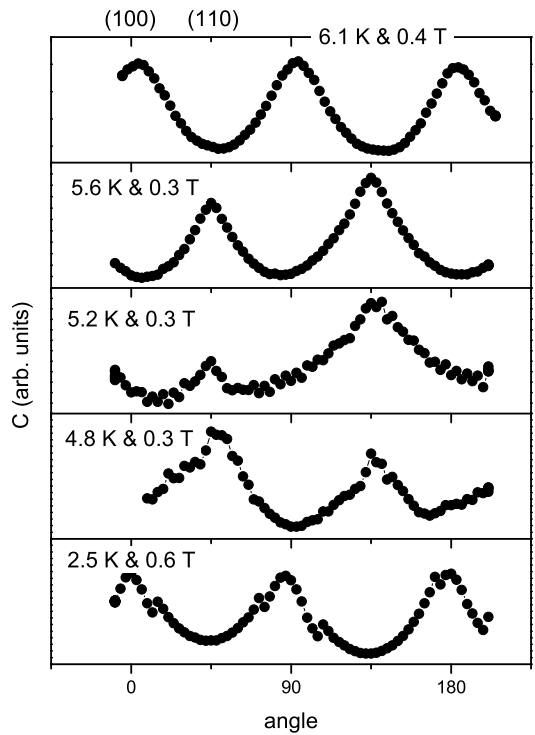}\centering
\caption{Heat capacity of sample B as a function of field angle at several
temperatures. The magnetic field is rotated within the basal plane and is
measured with respect to a-axis. The detailed condition is marked in each
responsible panel.}
\end{figure}

At 2.5 and 6.1~K, the low-field heat capacity has maxima for the field along 
$<100>$ directions. The area under the field-angle heat capacity is
proportional to magnetic entropy change, indicating that there is more
magnetic disorder for the field along $<100>$ than along $<110>$. The
field-angle entropy modulation can be explained by the fact that the net
moment of Ho$^{3+}$ ions is directed along $<110>$ directions in the
commensurate antiferromagnetic phase.\cite{lynn97} At 4.8, 5.2, and 5.6~K
where helical magnetic phases appear, the peak positions shifted by 45$%
^{\circ }$ to $<110>$ directions, indicating that the magnetic moments are
preferably aligned along $<100>$ directions. The peak intensities are also
asymmetric with field angle, which may be related to the oscillatory
magnetic structures observed\cite{goldman94,grigereit94,kreyssig97} in this temperature range.

Fig.~11 shows the field-angle heat capacity at 6.1~K in 0.5, 1, 2, and 4~T.
According to the magnetic phase diagram of HoNi$_{2}$B$_{2}$C (see Fig.~9),
the heat capacity should exhibit monotonic 4-fold angular oscillation at
high fields because there is no transition other than the $T_{M}$ transition
nearby this temperature. The peaks along [100] at 0.5~T, however, were split
into two peaks at 1~T and the minima along [110] were totally flattened out.
At 2~T, the distance between the two split peaks becomes narrower and the
flat minima become broad maxima. In 4~T, the split peaks merge and show a
delta function like peak and the broad maxima at [110] in 2~T returns to a
broad minima as was in 0.5~T. Two satellite peaks appear at $\pm 12^{\circ}$
of [100] peaks.
\begin{figure}[tbp]
\includegraphics[width=7cm,clip]{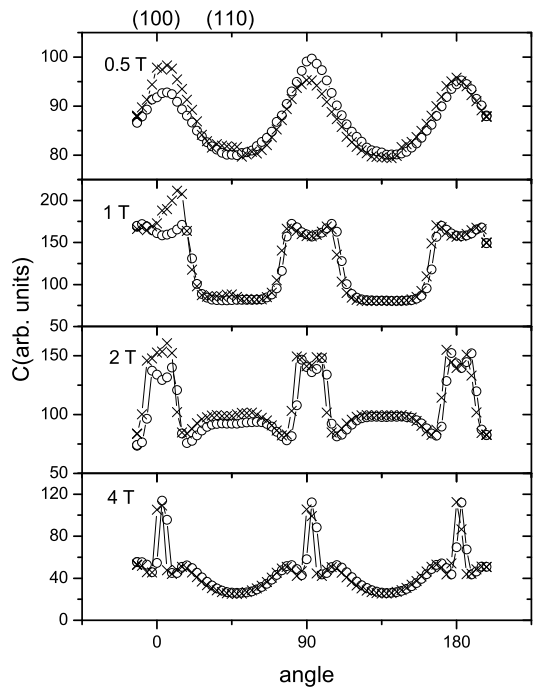}\centering
\caption{Heat capacity of sample B as a function of field angle at 6.1~K in
0.5, 1, 2, 4~T. The circles describe the data with rotating magnetic field
clockwise and the crosses, with rotating field counter clockwise.}
\end{figure}

The seemingly anomalous peak splittings along [100] may be explained as
following. In 1~T, $T_{M}$ moves from 3.7~K for $B\parallel \lbrack 110]$ to
6.5~K for $B\parallel \lbrack 100]$ or $T_{M}=3.7+0.031\alpha $ where $%
\alpha $ is the in-plane field angle measured against the $a$-axis. To move
the $T_{M}$ through 6.1~K means a field angle of 0.4~K/0.031~K$\cdot $deg$%
^{-1}\approx 12.5^{\circ }$, which is consistent with the peak positions at $%
10\pm 1.5$. At 2~T, the $T_{M}$ moves from 0~K for $B\parallel \lbrack 110]$
to 6.57~K for $B\parallel \lbrack 100]$, predicting 6.3$^{\circ }$ of peak
splitting. Experimentally, the observed splitting was $6\pm 1.5^{\circ }$.
The larger the field, the faster the peak has to move in angle. Even though
the anomalous peak splitting could be accounted for in terms of a simple
linear relation between $T_{M}$ and field angle $\alpha $, the interchange
between minima and maxima at [110] and the satellite peaks in 4~T have yet
to be understood, suggesting that the magnetic phase in high fields is not
as simple as was originally envisaged.

In Fig.~11, the circles describe the data when the in-plane magnetic field
was rotated clockwise and the crosses, when the field was rotated
counterclockwise. In this measurement, the angle spacing between two data
points is $3^{\circ}$. At 0.5~T, the two data sets do not show hysteresis in
phase, but show hysteresis in amplitude with field direction. Above 1~T,
there is a systematic shift in phase by $\sim 3^{\circ}$ with field
direction, indicating that the related transition may be a first order
magnetic transition. The experimental error bar in determining the field
angle is $\pm 1^{\circ}$, which is less than that of the angle shift in peak
positions. The hysteresis with field-angle direction may be another
manifestation of the first order nature of the metamagnetic transition from $%
(\uparrow \uparrow \rightarrow)$ to $(\uparrow \uparrow \uparrow)$ phases,
which were observed in neutron-diffraction study by Campbell et al.\cite%
{campbell00}

\section{Summary}

We have studied the magnetic superconductor HoNi$_{2}$B$_{2}$C via specific
heat as a function of temperature, magnetic field, and magnetic field angle.
The small value of the specific heat discontinuity at $T_{c}$ indicates that
Ho$^{3+}$ ions act as pair breakers as was predicted in Abrikosov-Gor'kov
theory. The temperature hysteresis of the zero-field specific heat at $T_{N}$
provides direct evidence that the magnetic AF transition is of first order,
accompanying a change from tetragonal to orthorhombic structure. The
similarity of $T^{\ast }$ transition to the $T_{N}$ transition indicates
that it is due to the $c^{\ast }$ spiral structure. The anomalous in-plane
anisotropy of $T_{M}$ transition was explained in terms of the $0.6~a^{\ast
} $ Fermi surface nesting feature.

This work at Urbana is supported by NSF DMR 99-72087. And the work at Pohang
was supported by the Ministry of Science and Technology of Korea through the
Creative Research Initiative Program. X-ray measurements were carried out in
the Center for Microanalysis of Materials, University of Illinois, which is
partially supported by the U.S Department of Energy under grant
DEFG02-91-ER45439. T. Park acknowledges benefits from the discussion with
Dr. J. D. Thompson.

\bibliography{hnbc}

\end{document}